\documentclass[onecolumn,12pt,nofootinbib,aps,a4paper]{revtex4}

\usepackage{graphicx}
\usepackage{bm}
\usepackage{dsfont}
\usepackage{mathrsfs} 
\usepackage{amssymb,amsfonts,amsmath}
\usepackage{graphicx}
\usepackage{color}
\usepackage{amsthm}

\newcommand{\dd}{\mathrm{d}}
\newcommand{\ee}{\mathrm{e}}
\newcommand{\ii}{\mathrm{i}}
\newcommand{\R}{\mathds R}
\newcommand{\C}{\mathds C}
\newcommand{\eps}{\varepsilon}

\newcommand{\E}{\mathcal E}

\newcommand{\Sth}{\ensuremath{\mathbb S^3}}
\newcommand{\Al}{\mathcal A_1}
\newcommand{\Ar}{\mathcal A_2}
\newcommand{\Hp}{{\mathcal H_\mathrm{p}}}
\newcommand{\Hf}{{\mathcal H_\mathrm{f}}}
\newcommand{\Sp}{{\mathcal S_\mathrm{p}}}
\newcommand{\Sf}{{\mathcal S_\mathrm{f}}}
\newcommand{\p}{_\mathrm{p}}
\newcommand{\f}{_\mathrm{f}}

\def\Xint#1{\mathchoice
   {\XXint\displaystyle\textstyle{#1}}%
   {\XXint\textstyle\scriptstyle{#1}}%
   {\XXint\scriptstyle\scriptscriptstyle{#1}}%
   {\XXint\scriptscriptstyle\scriptscriptstyle{#1}}%
   \!\int}
\def\XXint#1#2#3{{\setbox0=\hbox{$#1{#2#3}{\int}$}
     \vcenter{\hbox{$#2#3$}}\kern-.5\wd0}}

\def\dashint{\Xint-}

\normalsize

\theoremstyle{plain}

\begin{document}

\title{New Gowdy-symmetric vacuum and electrovacuum solutions}

\author{\bf J\"org Hennig}
\email{jhennig@maths.otago.ac.nz}

\affiliation{Department of Mathematics and Statistics,
           University of Otago,
           P.O. Box 56, Dunedin 9054, New Zealand}

\begin{abstract}
We construct a 4-parameter family of inhomogeneous cosmological models, which contains two recently derived 3-parameter families as special cases. The corresponding exact vacuum solution to Einstein's field equations is obtained with methods from soliton theory. We also study properties of these models and find that they combine all interesting features of both earlier solution families: general regularity within the maximal globally hyperbolic region, particular singular cases in which a curvature singularity with a directional behaviour forms, a highly non-trivial causal structure, and Cauchy horizons whose null generators can have both closed or non-closed orbits. In the second part of the paper, we discuss the generalisation from vacuum to electrovacuum. Moreover, we also present a family of exact solutions for that case and study its properties.
\end{abstract}

\maketitle

\mbox{}\vspace{-1.5cm}\mbox{}
\section{Introduction}

General relativity is one of the most accurate and mathematically elegant theories of modern physics. Quite appropriately, in his classification of physical theories as 1. SUPERB, 2. USEFUL, or 3. TENTATIVE, Roger Penrose \cite{Penrose89} clearly places it in the first category. In the context of cosmology, it was the investigation of solutions to Einstein's field equations and their comparison with observations that  led to the detailed picture of our universe and its evolution that we currently have. Astonishing implications like the existence of dark energy and dark matter, the accelerated expansion and the indication that everything arose out of a big bang are all part of general relativistic cosmological models. Moreover, no alternative theory of gravity (like string theory, loop quantum gravity, Brans-Dicke theory, etc.) has yet provided a model of the large-scale universe that makes experimentally testable predictions that are beyond those of general relativity.

On the other hand, it is also well-known that not \emph{all} mathematical solutions to the Einstein equations are physically acceptable. Some cosmological models violate reasonable assumptions that are widely believed to hold in the real universe, like causality and determinism. A famous example of a solution with such an interesting pathological behaviour is the Taub solution \cite{Taub51}, a 2-parameter family of spatially homogeneous cosmological models. These maximal globally hyperbolic spacetimes \cite{Choquet1969} are perfectly well-behaved in the time interval in which they are initially defined. However, they can be extended into the future and past to obtain the Taub-NUT solutions \cite{Misner1963, MisnerTaub1969, NUT1963}. Interestingly, there are several non-equivalent extensions \cite{ChruscielIsenberg1993}, so that information about the solutions in the original domain does not uniquely determine which of these extensions is realised --- determinism breaks down. Moreover, beyond the Cauchy horizons (the hypersurfaces that separate the maximal globally hyperbolic region from the extended regions) there exist closed causal curves, which would allow observers or photons to travel back in time --- clearly a violation of causality.

These unusual properties of the Taub-NUT solution have led to the question as to whether such a pathological behaviour is a typical feature of solutions to Einstein's field equations, or whether it can only occur under very special circumstances, such as for spacetimes with a high degree of symmetry or very restricted types of matter. According to the famous strong cosmic censorship conjecture \cite{Penrose69, MoncriefEardley1981, Chrusciel1991} (see also \cite{Rendall2005, Ringstrom2009, Isenberg2015} for more details and further references), the maximal globally hyperbolic development of generic data on some Cauchy surface is \emph{inextendible}. If this hypothesis were true, it would imply that solutions  which, like the Taub solution, can be extended into acausal regions are rare exceptions. They could only occur for a (in some sense) negligibly small set of ``non-generic'' initial data. Unfortunately, in the current state of the art, it seems unrealistic to prove or disprove the strong cosmic censorship conjecture in full generality. Instead, a practicable and promising approach is to start from suitable small classes of symmetric models and investigate their properties. Then one can try to gradually increase the complexity of the models and explore their behaviour. For each class of models, one can attempt to prove existence of solutions with Cauchy horizons and acausal regions, and study properties like the presence of curvature singularities.

Along those lines, the existence of a class of vacuum solutions with Cauchy horizons --- the \emph{generalised Taub-NUT spacetimes} --- was shown by Moncrief \cite{Moncrief1984}. More recently, a different class of cosmological models with Cauchy horizons has been introduced: the  \emph{smooth Gowdy-symmetric generalised Taub-NUT (SGGTN) solutions}. Existence of these inhomogeneous cosmological vacuum models was shown in \cite{BeyerHennig2012, Hennig2014}. Compared to Moncrief's class of solutions, the SGGTN models require less regularity (smoothness instead of analyticity) but more symmetry (Gowdy symmetry with two spacelike Killing vectors instead of just one). A number of properties of these solutions can be studied with abstract methods, but in order to get a more complete picture of the behaviour of some of these models, it is desirable to obtain families of exact solutions. 

Recently, two families of exact SGGTN solutions were constructed with methods from soliton theory \cite{BeyerHennig2014,Hennig2014}. Both families contain Cauchy horizons and acausal regions, but their causal and singularity structure is quite different, as will be summarised below. The goal of the present paper is to construct more general families of exact solutions, again using methods from soliton theory. The nature of the desired generalizations is twofold: (i) we want to embed the two above-mentioned exact solution families into a larger family that contains both as special cases, a family which combines all of the interesting properties that these two families have, and (ii) to extend the considerations from vacuum to electrovacuum, in order to include a non-trivial additional field, and to construct an exact solution family for that case too. 

To this end, we will start by summarizing the definition and important properties of the class of SGGTN solutions in Sec.~\ref{sec:SGGTN}. In the same section, we also give an overview of the two exact solution families that we intend to generalize. In Sec.~\ref{sec:vacuum} we present a new family of vacuum solutions and discuss their properties. Afterwards, in Sec.~\ref{sec:newelectrovac}, we work out the generalization of the underlying equations to electrovacuum and derive a corresponding family of exact solutions. Finally, in Sec.~\ref{sec:discussion}, we discuss our results.

\section{Smooth Gowdy-symmetric generalised Taub-NUT solutions}\label{sec:SGGTN}
\subsection{General setup}

In this subsection we summarize some properties of the ``smooth Gowdy-symmetric generalised Taub-NUT (SGGTN) solutions''. For more details we refer to \cite{BeyerHennig2012,Hennig2014}. 

The SGGTN models are solutions to the Einstein vacuum equations, their spatial topology is $\Sth$ --- the three-sphere topology ---  and they have a past and (generally) a future Cauchy horizon. In terms of a time coordinate $t$ and spatial coordinates $\theta$, $\rho_1$, $\rho_2$ (which are adapted to the two Killing vectors $\partial_{\rho_1}$ and $\partial_{\rho_2}$), the line element is given by
\begin{equation}\label{eq:metric0}
  \dd s^2=\ee^M(-\dd t^2+\dd\theta^2)+R_0\left[\sin^2\!t\,\ee^u (\dd\rho_1+Q \dd\rho_2)^2+\sin^2\!\theta\,\ee^{-u} \dd\rho_2^2\right].
\end{equation}
Here $u$, $Q$ and $M$ are functions of $t$ and $\theta$ alone, and $R_0$ is a positive constant. The angles $\rho_1$, $\rho_2$ take on values in the regions
\begin{equation}\label{eq:rhodomain}
 \frac{\rho_1+\rho_2}{2}\in(0,2\pi),\quad
 \frac{\rho_1-\rho_2}{2}\in(0,2\pi),
\end{equation}
and the solutions are initially defined in the ``Gowdy square'' with
\begin{equation}
 t\in(0,\pi),\quad \theta\in(0,\pi).
\end{equation}
The boundaries $\theta=0,\pi$ of the Gowdy square are symmetry axes, and the boundaries $t=0,\pi$ correspond to the past and future Cauchy horizons $\Hp$ and $\Hf$. Moreover, the solutions can be extended through these horizons.

We do not list the corresponding Einstein vacuum equations here, since they can be obtained from the electrovacuum equations \eqref{eq:ueq}--\eqref{eq:Meq2} in Appendix~\ref{sec:EM} in the limit of vanishing 4-potential ($A_3=A_4=0$). However, we note that the essential part of the field equations is equivalent to the Ernst equation
\begin{equation}\label{eq:EE}
 \Re(\E)\cdot \left(-\E_{,tt}-\cot t\,\E_{,t}
        +\E_{,\theta\theta}+\cot\theta\,\E_{,\theta}\right)
 =-\E_{,t}^{\ 2}+\E_{,\theta}^{\ 2}
\end{equation} 
for the complex Ernst potential $\E=f+\ii b$.

The SGGTN solutions have two functional degrees of freedom, namely the functions $S_{**}(\theta)$ and $Q_{*}(\theta)$ considered in \cite{BeyerHennig2012} (these are called ``asymptotic data'' and describe the behaviour of the solution as the past Cauchy horizon $\Hp$ is approached), or, equivalently, the initial Ernst potential at the past Cauchy horizon, $\E(t=0,\theta)$. In the original setup in \cite{BeyerHennig2012}, the initial data have to be chosen subject to a periodicity condition. This condition ensures that the past horizon is generated by the Killing vector $\partial_{\rho_1}$, which has \emph{closed} orbits. In this situation, the metric functions satisfy the following regularity conditions at the axes $\Al\ (\theta=0)$ and $\Ar\ (\theta=\pi)$,
\begin{equation}\label{eq:regcond}
 \Al: \quad Q=1,\quad \ee^{M+u}=R_0,\qquad
 \Ar: \quad Q=-1,\quad \ee^{M+u}=R_0.
\end{equation}

In \cite{Hennig2014} this was generalised to SGGTN spacetimes with past horizons that are ruled by \emph{non-closed} null-generators. 
Investigations of Cauchy horizons with this property are particularly interesting as some theorems about cosmological spacetimes assume closedness of the generators \cite{FriedrichRaczWald1999,MoncriefIsenberg1983,Racz2000}. Hence it is useful to have examples of cosmological models that violate this assumption.
In the case of non-closed generators, no periodicity condition is required for the asymptotic data, and the past horizon is generated by the null vector 
\begin{equation}\label{eq:ap}
 \partial_{\rho_1}-a\p\partial_{\rho_2}, 
\end{equation}
where $a\p=\textrm{constant}$. Also in this more general setup, the metric potentials still have to satisfy the regularity conditions \eqref{eq:regcond}. With an appropriate change of the Killing basis (corresponding to a linear transformation of the coordinates $\rho_1$, $\rho_2$) it is then possible to arrange that the past horizon is again generated by $\partial_{\rho_1}$ (in terms of a new coordinate that we again denote by $\rho_1$). This, however, changes the boundary conditions of the transformed metric potentials and the new constant $R_0$ to
\begin{equation}\label{eq:regcond1}
 \Al: \quad Q=1,\quad \ee^{M+u}=\frac{R_0}{(1+a\p)^2},\qquad
 \Ar: \quad Q=-1,\quad \ee^{M+u}=\frac{R_0}{(1-a\p)^2}.
\end{equation}
Note that the new coordinates $\rho_1$, $\rho_2$ are no longer defined in the region \eqref{eq:rhodomain}, but in a different domain (cf.\ \cite{Hennig2014}, and see the  related discussion in Appendix \ref{sec:axiscoords} here). However, the form of the metric \eqref{eq:metric0} is unchanged. In the following we will exclusively use these new coordinates and potentials.

Note that the solutions are regular in the entire Gowdy square, but potentially singular at the future boundary $t=\pi$. This occurs if the imaginary part $b$ of the initial Ernst potential satisfies the condition
\begin{equation}\label{eq:singcond}
 b_B-b_A=\pm4,
\end{equation}
where $b$ is evaluated at the points
\begin{equation}\label{eq:AB}
 A\,(t=0,\theta=0)\quad\textrm{and}\quad B\,(t=0,\theta=\pi). 
\end{equation}
In these ``singular cases'' the solutions have a curvature singularity at $t=\pi$, $\theta=0$ (for a `$+$' sign) or at $t=\pi$, $\theta=\pi$ (for a `$-$' sign). With the condition \eqref{eq:singcond} we can therefore already read off from the initial data at $t=0$ whether or not the solution will develop a singularity at $t=\pi$.

Finally, we note that the solutions in the extended regions beyond the Cauchy horizons can be regular or singular, depending on the particular solution, and we will discuss several examples below.

In the next two subsections we give a short summary of the two exact SGGTN solutions that were constructed in \cite{BeyerHennig2014,Hennig2014}.
\subsection{``Solution 1''}\label{sec:sol1}
An SGGTN solution for which the past Cauchy horizon has closed null generators, corresponding to $a\p=0$ in \eqref{eq:regcond1}, can be obtained by solving an initial value problem for the Ernst equation with appropriate initial data at $t=0$. Generally, the real part $f$ and the imaginary part $b$ of the initial Ernst potential, considered as functions of
\begin{equation}
 x:=\cos\theta,
\end{equation}
have to be chosen subject to the following conditions, which are necessary to guarantee existence of a corresponding regular metric:
\begin{equation}\label{eq:Ernstcond}
 f_A=0,\quad f_B=0,\quad
 f_{,x}|_A=-\left|\frac{1+a\p}{1-a\p}\right|f_{,x}|_B,\quad
 b_{,x}|_A=-2,\quad b_{,x}|_B=2.
\end{equation}
Here, the functions are again evaluated at the above-mentioned points $A$ and $B$, cf.\ \eqref{eq:AB}.
Consequently, a solution with $a\p=0$ is obtained by choosing $f$ such that $f_{,x}|_A=-f_{,x}|_B$. An initial Ernst potential subject to these conditions was studied in \cite{BeyerHennig2014}, namely
\begin{equation}\label{eq:indat1}
 t=0:\quad \E=c_1(1-x^2)
      +\ii x\left[c_3(x^2-3)-x\right],
\end{equation}
where the real part is quadratic and the imaginary part cubic in $x$. 
For the derivation of the corresponding solution we refer to \cite{BeyerHennig2014}. The final results are the metric potentials in terms of $x$ and 
\begin{equation}
 y:=\cos t, 
\end{equation}
which are given by the following expressions,
\begin{eqnarray}\label{eq:solnew1}
 \ee^M & = & \frac{R_0}{64 c_1^3}(U^2+V^2),\quad
 \ee^u = \frac{16c_1U}{(1+y)(U^2+V^2)},\\
 \label{eq:solnew2}
 Q & = & x+\frac{c_3}{8}(1-x^2)\left(7+4y+y^2+\frac{(1-y)V^2}{4c_1^2U}\right),
\end{eqnarray}
where
\begin{equation}\label{eq:defUV}
 U  := c_3^2(1-x^2)(1-y)^3+4c_1^2(1+y),\quad
 V  :=  4c_1(1-y)[1-c_3x(2+y)].
\end{equation}
This family of solutions depends on the three parameters $c_1>0$, $c_3\in\R$ and $R_0>0$.

In the special case $c_3=0$, the solution reduces to the spatially homogeneous Taub solution, and for $c_3\neq0$ we obtain more complicated inhomogeneous solutions.

According to the singularity condition \eqref{eq:singcond}, we expect singularities for $c_3=\pm1$, but otherwise the solutions are regular in the Gowdy square and at the boundaries. This can be readily verified from the explicit solution. Indeed, for $c_3=\pm 1$, the Kretschmann scalar $K=R_{ijkl}R^{ijkl}$ diverges at the singular point. Interestingly, this divergence is of a directional nature: as the singularity at $t=\pi$, $\theta=0$ or $t=\pi$, $\theta=\pi$ is approached, $K$ either tends to $+\infty$, or to $-\infty$, or even to any finite value, depending on the path along which the singular point is approached. This is similar to the behaviour of the Curzon solution \cite{Curzon1925}, as was discussed in \cite{BeyerHennig2014}. However, whereas the Curzon solution allows for extensions through its so-called \emph{directional singularity} \cite{ScottSzekeres1986a,ScottSzekeres1986b}, the same is not possible for ``solution 1''. This follows from the observation that along each curve towards the singularity on which $K$ remains finite, other curvature invariants diverge. Hence the spacetime geometry is still singular along the paths with regular $K$.
Therefore, ``solution 1'' is an interesting example of a solution where the Kretschmann scalar alone does not contain the complete information about the singularity structure of the spacetime. 

Extensions of the solution through the Cauchy horizons $\Hp$ and $\Hf$ can be constructed by reexpressing the line element in terms of the coordinates $x$ and $y$ instead of $t$ and $\theta$ (and after an additional transformation of $\rho_1$ and $\rho_2$). The solution is initially defined in the Gowdy square, corresponding to $(x,y)\in(-1,1)\times(-1,1)$, but it can be extended to all $y\in\R$. It turns out that there are several possible extensions, and they all contain curvature singularities (one singularity in the past extensions, and either one or two singularities in the future extensions, depending on the parameter values). 

Finally, we note that ``solution 1'' was recently reproduced numerically, in order to test the numerical accuracy of a new evolution code for cosmological spacetimes with spatial $\Sth$ topology and $U(1)$ symmetry \cite{Beyer2015}. 

\subsection{``Solution 2''}\label{sec:sol2}

An SGGTN solution with $a\p\neq0$ was derived in \cite{Hennig2014}. This was obtained from the initial data
\begin{equation}\label{eq:indat2}
t=0:\quad \E =c_1(1-x^2)\left(1-\frac{x}{d}\right)-\ii x^2,
\end{equation}
which has a cubic real part and a quadratic imaginary part. The corresponding solution reads
\begin{eqnarray}
 \ee^u &=& 8 c_1 d
       \frac{ 4d^2(d+r_d-xy)+c_1^2(1-x^2)^2(1+y)^4/(d+r_d-xy)}
       {\big[4d^2(1-y)+c_1^2(1-x^2)(1+y)^3\big]^2+\big[4c_1 d(1+y) r_d\big]^2},\\
  Q &=& x+\frac12(d-r_d-xy),\\
  \ee^M &=&
      \frac{dR_0}{16 c_1(1-d^2)^2r_d}\left[4d^2(r_d + x - dy)^2 \right.\\ 
     &&\quad\left.  +c_1^2 (1 + y)^2 \Big(r_d^2 + (d x - y) (1 + y) + (d + x) r_d\Big)^2\right],
\end{eqnarray}
where
\begin{equation}
  r_d = \sqrt{(d-xy)^2-(1-x^2)(1-y^2)}.
\end{equation}
The three parameters are $R_0>0$, $c_1>0$ and $d>1$ (where $d$ determines the value of $a\p$, see Eq.~\eqref{eq:dap} below).

For this solution we have $b_B-b_A=0$ for all parameter values. Hence the singularity condition \eqref{eq:singcond} is never satisfied and the spacetime is always regular in the Gowdy square. 

Also for this solution one can introduce suitable coordinates in which extensions through the Cauchy horizons can be constructed. It turns out that the resulting causal structure is more complex than the causal structure of ``solution 1''. In particular, there are further horizons in the extended regions through which the solution can be extended too. For details we refer to \cite{Hennig2014} or to Sec.~\ref{sec:ext} below, where our new and more general solution is discussed, and turns out to have the same causal structure as this solution. (The explicit form of the coordinate transformations for the solution below is just slightly more complicated than the transformations needed for ``solution 2'' here.)

Finally, it turns out that the solution is everywhere free of singularities --- not only in the Gowdy square, but also in all extensions.

\section{A new family of vacuum solutions}\label{sec:vacuum}

In the following we intend to find a more general 4-parameter family of solutions that contains the 3-parameter solutions 1 and 2 described above as special cases.
\subsection{Initial data}

We intend to solve the Ernst equation for the initial Ernst potential 
\begin{equation}\label{eq:indat}
 t=0:\quad \E=c_1 (1-x^2)\left(1 - \frac{x}{d}\right) 
             + \ii x \left[c_3(x^2-3)-x)\right].
\end{equation}
Here both the real and imaginary parts of $\E$ are cubic functions of $x$. Since  \eqref{eq:indat} reduces to the initial data \eqref{eq:indat1} of ``solution 1'' for $d\to\infty$, and to the data \eqref{eq:indat2} for ``solution 2'' as $c_3\to0$, the corresponding solution will contain both of these earlier solutions as special cases. 

Before we solve this problem, we can already determine the parameter values that give singularities. We find
\begin{equation}
 b_B-b_A=4c_3,
\end{equation}
so that the condition $b_B-b_A=\pm4$ [cf.\ \eqref{eq:singcond}] is satisfied for $c_3=\pm1$. Consequently, singularities are expected if and only if $c_3=\pm1$ (exactly as for ``solution 1'').

A physical interpretation of the constant $d$ is provided by its relation to the generators of the past Cauchy horizon $\Hp$. $d$ turns out to be completely fixed in terms of the parameter $a\p$, which was introduced in \eqref{eq:ap}. To this end we note that the integrability condition for the $M$-equations leads to the above-mentioned condition [cf.~\eqref{eq:Ernstcond}]
\begin{equation}
 f_{,x}|_A=-\left|\frac{1+a\p}{1-a\p}\right|f_{,x}|_B
\end{equation}
for the initial Ernst potential $\mathcal E=f+\ii b$, i.e.\ only for initial data subject to this condition does a regular solution $M$ exist. Without loss of generality, we can choose\footnote{Note that an ``inversion'' (an interchange of the Killing fields, corresponding to a coordinate transformation $\rho_1\mapsto\rho_2$, $\rho_2\mapsto\rho_1$) leads to $a\p\mapsto 1/a\p$. Hence we can always achieve $|a\p|<1$. Moreover, the coordinate transformation $\rho_2\mapsto-\rho_2$ leads to $a\p\mapsto-a\p$, which allows us to fix the sign of $a\p$. We will choose $a\p$ to be negative (since the parameter $d$ will then be positive).} $a\p$ in the interval $-1<a\p\le 0$, which results in
\begin{equation}\label{eq:dap}
 d=-\frac{1}{a\p}.
\end{equation}
This shows that $d>1$ and that the limit $a\p\to0$ corresponds to $d\to\infty$.
\subsection{Solution of the initial value problem}

In order to solve an initial value problem for the Ernst equation with initial data \eqref{eq:indat}, we use the integral equation method due to Sibgatullin \cite{Sibgatullin}. Since this procedure has been described in detail in \cite{BeyerHennig2014}, we only give a short summary here.

``Sibgatullin's integral method'' was originally developed to construct \emph{axisymetric and stationary} spacetimes (with a spacelike and a \emph{timelike} Killing vector) and not Gowdy-symmetric solutions. Nevertheless, with the formal, complex coordinate transformation 
$$\rho  = \ii \sin t\sin\theta,\quad \zeta = \cos t\cos\theta$$
to coordinates ($\rho,\zeta,\rho_1,\rho_2$), our metric \eqref{eq:metric0} takes the Weyl-Lewis-Papapetrou form for axisymmetric and stationary spacetimes. Hence we have translated our initial value problem into a problem that can be tackled with Sibgatullin's method.

The idea of this method is the following.
For a given initial Ernst potential at $t=0$ (corresponding to $\rho=0$), we consider the analytic continuation
$e(\xi):=\E(\rho=0,\zeta=\xi)$, $\tilde e(\xi):=\overline{e(\bar\xi)}$ with $\xi\in\C$.
Then we solve the integral equation
\begin{equation}\label{eq:inteq}
 \dashint_{-1}^1\frac{\mu(\xi;\rho,\zeta)[e(\xi)+\tilde e(\eta)]\,\dd\sigma}{(\sigma-\tau)\sqrt{1-\sigma^2}}=0
\end{equation}
for $\mu(\xi;\rho,\zeta)$ subject to the constraint
\begin{equation}\label{eq:constraint}
 \int_{-1}^1\frac{\mu(\xi;\rho,\zeta)\,\dd\sigma}{\sqrt{1-\sigma^2}}=\pi.
\end{equation}
Here, $\dashint$ denotes the principal value integral and $\xi:=\zeta+\ii\rho\sigma$, $\eta:=\zeta+\ii\rho\tau$ with $\sigma,\tau\in[-1,1]$. 
If the function $e(\xi)$ is a rational function in $\xi$ (as in the present situation), the solution $\mu$ will be a rational function as well, and with the  structure of $\mu$ being known, the problem can be reduced to a system of algebraic equations for the unknown coefficients in $\mu$, which are functions of $\rho$ and $\zeta$.

Once the integral equation is solved, the corresponding Ernst potential can be obtained from another integration,
\begin{equation}\label{eq:EP}
 \E(\rho,\zeta)=\frac{1}{\pi}\int_{-1}^1\frac{e(\xi)\mu(\xi)\dd\sigma}{\sqrt{1-\sigma^2}}.
\end{equation}
From $\E$ we can construct the metric potentials $u$, $Q$ and $M$ with a combination of algebraic manipulations and line integrations (or, equivalently, by solving first-order PDEs), see \cite{BeyerHennig2014,Hennig2014} for details. This computation is somewhat involved and we found that only with an appropriate combination of Maple and Mathematica all required integrals could be computed.\footnote{An example is the computation of the metric potential $M$. After $u$ and $Q$ are available, one finds explicit expressions for $M_{,x}$ and $M_{,y}$. The attempt to integrate $M_{,x}$ with respect to $x$ in Mathematica, in the special case $c_3=0$ (i.e.\ the case of ``solution 2'') leads to a complicated expression that involves the sum over roots of a fourth-order polynomial. With Maple, the same integration can be performed without problems and results in a much simpler expression. If we try to integrate $M_{,x}$ with Mathematica in the general case ($c_3\neq 0$), then the computation does not stop at all within a few hours, after which my computer crashed. With Maple the computation could again be done without difficulties. On the other hand, Mathematica turned out to provide more efficient simplifications of the lengthy expressions than Maple.}
Moreover, some of the intermediate results turn out to be extremely lengthy (Mathematica's \verb'ByteCount' function, which gives the number of bytes that are used to store a term, returned more than 15 million for some of these) and the computations and simplifications required several hours. Luckily, the final results are surprisingly simple and can be given in a relatively compact form, if appropriate abbreviations are used for terms that appear repeatedly. 

The final metric potentials, in terms of parameters $R_0>0$, $c_1>0$, $c_3\in\R$ and $d>1$, are the following,
\begin{eqnarray}
 \ee^u &=&  8 c_1 d\frac{U}{(1 + y) (U^2 + V^2)}\label{eq:formulau},\\
 Q     &=&  x + \frac{1-x^2}{4}
           \Bigg[2 c_3(2 + y)\nonumber\\
       && \qquad 
        + \frac{1}{d+r_d-xy} \left(\frac{(1 + y)V}{c_1 d} 
                + \frac{c_3 (1-y)V^2}{2c_1^2U} - c_3 r_d(1-y^2)\right)\Bigg],\label{eq:formulaQ}\\
 \ee^M &=& c\,\frac{W^2}{r_d}\,(U^2 + V^2),\label{eq:formulaM}
\end{eqnarray}
where (as before) $x=\cos\theta$, $y=\cos t$ and
\begin{eqnarray}
 r_d &=& \sqrt{(d-xy)^2-(1-x^2)(1-y^2)},\label{eq:defrd}\\
 U &=& c_1^2 (d+r_d-xy)(1+y) 
      + c_3^2d^2\frac{(1-x^2)(1 - y)^3}{d+r_d-xy},\label{eq:defU1}\\
   & \equiv & \frac{1}{1+y}\left[c_1^2
(d+r_d-xy)(1+y)^2+c_3^2d^2(d-r_d-xy)(1-y)^2\right],\label{eq:defU2}\\
 V &=& 2 c_1 d [1-c_3(d-r_d+2x)](1-y),\label{eq:defV}\\
 W &=& \frac{r_d+x-dy}{1-y}\label{eq:defW1}\\
   &\equiv& d+x - \frac{(1-x^2)(1+y)}{d+r_d-xy}\label{eq:defW2}\\
   &\equiv& \frac{\sqrt{(dy-x)^2 + (d^2-1)(1-y^2)}-(dy-x)}{1-y}.\label{eq:defW3}
\end{eqnarray}
(Note that the first formula for $W$ gives a concise expression, but might indicate a singularity at $y=1$. However, it follows from the second formulation that $W$ is actually regular there, for $d+r_d-x y$ is always positive at $\Hp$. The third expression shows that $W$ is nonnegative in the Gowdy square, even though potentially zero at $y=1$. But evaluating the second formulation at $y=1$ shows that $W=(d^2-1)/(d-x)>0$ there.)

The integration constant $c$ in the formula for $M$ is fixed by the boundary conditions on the axes.\footnote{Note that we have two axes conditions, but only one constant $c$. However, as expected, the conditions at both axes leads to the same value for $c$, since the initial data satisfy the constraint that ensures integrability of the $M$-equations.} We obtain (for our choice $|a\p|<1$)
\begin{equation}
 c=\frac{d R_0}{16 c_1^3 (d^2 - 1)^2}.
\end{equation}

Finally, one can easily check that for $a\p\to 0$ (i.e. $d=-1/a\p\to\infty$) this solution reduces to ``solution 1'', whereas for $c_3\to 0$ we obtain ``solution 2''.

\subsection{Regularity in the Gowdy square}
In the discussion of ``solution 1'' in \cite{BeyerHennig2014} it was observed that the Kretschmann scalar is proportional to $\ee^{-6M}$, and the solution was regular wherever $\ee^M\neq0$ holds. The same applies to the present solution, see Eq.~\eqref{eq:Kret} below. Hence we first also consider zeros of $\ee^M$ here. 

Since $W$ and $r_d$ are strictly positive in the Gowdy square, $\ee^M$ can only vanish at simultaneous zeros of $U$ and $V$. The two terms in $U$ are both nonnegative, so that both must vanish at a zero of $U$. The first term vanishes at $y=-1$ and the second term at $x=\pm 1$. Therefore, $U$ has the two zeros $x=\pm1$, $y=-1$. For this choice of $x$ and $y$, $V$ becomes 
$4c_1 d (1\mp c_3)$, which vanishes if and only if $c_3=\pm 1$ holds. Hence there are only two values of $c_3$ for which $\ee^M$ has zeros,
\begin{equation}
 c_3=\pm1:\quad \ee^M=0\quad\textrm{at}\quad
 x=\pm 1,\ y=-1.
\end{equation}
This corresponds exactly to the positions where singularities were expected in the  singular cases $c_3=\pm1$ mentioned previously.

Similarly, $\ee^u$ is regular and positive in the Gowdy square with exception of the future horizon $\Hf$, where it diverges (unless $c_3=0$, in which case $U$ is proportional to $(1+y)$ and thus compensates for the $1/(1+y)$-term in the formula for $\ee^u$). This divergence, however, is just due to the particular parametrization of the metric in terms of the functions $u$, $M$ and $Q$ that we have chosen and does not indicate any physical irregularity of the spacetime. Exactly the same behaviour was also observed for ``solution 1''.

Finally, $Q$ is regular except for discontinuities at the points $C\ (\theta=0,\,t=\pi)$ and $D\ (\theta=t=\pi)$. $Q$ has the following values  on the boundary at the axes and the horizons,
\begin{eqnarray}
 \mathcal A_{1/2}: && Q=\pm1,\\
 \Hp:              && Q= x+\frac32 c_3 (1-x^2),\\
 \Hf:              && Q=\frac{1 + c_3^2}{2 c_3}.
\end{eqnarray}
Note that generally for the SGGTN spacetimes (i.e.\ not just for this particular family of solutions), the potential $Q$ is discontinuous at the boundary points on the horizons that are generated by a linear combination of both Killing vectors. However, one can make the discontinuity disappear by choosing an adapted Killing basis. On the other hand, $Q$ would be continuous at $C$ or $D$ if $c_3=\pm1$ holds, i.e.\ in the singular cases. So we see that a discontinuity in our Killing basis is even necessary for physical regularity.
\subsection{Kretschmann scalar}

As mentioned in Sec.~\ref{sec:sol1},  ``solution 1'' shows a directional behaviour of the Kretschmann scalar near the singularity in the singular cases. Here we study the corresponding situation for the generalised solution under discussion and investigate the behaviour of the Kretschmann scalar $K=R_{ijkl}R^{ijkl}$ near the point $C$  ($\theta=0$, $t=\pi$), where the solution becomes singular for $c_3=1$.

Note that for the nonsingular case $c_3\neq 1$, an expansion in polar coordinates centred at $C$,
\begin{equation}\label{eq:polar}
 x=1-r\cos\phi,\quad
 y=-1+r\sin\phi,\quad
 r\ge0,\quad
 \phi\in\left[0,\frac{\pi}{2}\right],
\end{equation}
leads to
\begin{equation}\label{eq:KretReg}
 K=\frac{12 (d-1)^4 \Big(c_1^2 d^2 [d-1+c_3(5+3d)]^2 
    -\left[4c_3^2 d^2 - c_1^2(d+1)^2\right]^2\Big)}{(c_3-1)^6 d^8 R_0^2}
  +\mathcal O(r).
\end{equation}
Evidently, $K$ is regular at $C$ and approaches the same constant value along any path towards $C$, i.e.\ for $r\to0$.

In the singular case $c_3=1$ [where \eqref{eq:KretReg} is not defined], expressed in the same polar coordinates,
we have the following leading-order behaviour of the Kretschmann scalar,
\begin{equation}
 K=\frac{g(\phi)}{r^6}+\mathcal O(r^{-5}),
\end{equation}
where
\begin{equation}
 g(\phi)=  \frac{768c_1^6d^6(d^2-1)^4 (d+1)^2 (1+T^2)^3}
    {R_0^2\left[4d^2+c_1^2(d+1)^2\right]^2 
     \left[4d^2+c_1^2 (d+1)^2 T^2\right]^6}\,
   p_+\left(\frac{c_1(d + 1)}{d} T\right) 
   p_-\left(\frac{c_1(d + 1)}{d} T\right),
\end{equation}
\begin{equation}
 T:=\tan\phi\in[0,\infty], 
\end{equation}
\begin{equation}
 p_{\pm}(x):= \pm c_{\pm} x^3 + 6 c_{\mp}x^2 \mp 12 c_{\pm} x - 8 c_{\mp},
 \quad
 c_{\pm} := 2 \pm \frac{c_1(d+1)}{d}.
\end{equation}
Note that $c_+$ is always positive, whereas $c_-$ can be positive, zero or negative. It easily follows from the structure of the polynomials $p_\pm$ that, for $c_->0$, $p_+$ has one nonnegative zero\footnote{Due to the requirement $T\ge 0$, we are only interested in nonnegative arguments of $p_{\pm}$.} and $p_-$ has two nonnegative zeros. For $c_-\le0$ the situation is reversed. Furthermore, there cannot be simultaneous zeros of $p_+$ and $p_-$, i.e.\ all zeros are distinct. Hence the product $p_+(x)p_-(x)$ always has three nonnegative zeros, corresponding to three $\phi$-directions along which the leading order term of $K$ vanishes. In the case $c_-=0$, i.e.\ for $c_1=2d/(d+1)$, the function $g(\phi)$ has an additional zero at $\phi=\pi/2$ (corresponding to $T\to\infty$). Since $g(\phi)$ changes sign at each of these (simple) zeros, $K$ can approach both $+\infty$ and $-\infty$, depending on the direction in which the singular point is approached. This is exactly the behaviour that ``solution 1'' exhibits, i.e.\ these properties carry over to the more general solution here.

\subsection{Extensions}\label{sec:ext}

\begin{figure}\centering
 \includegraphics{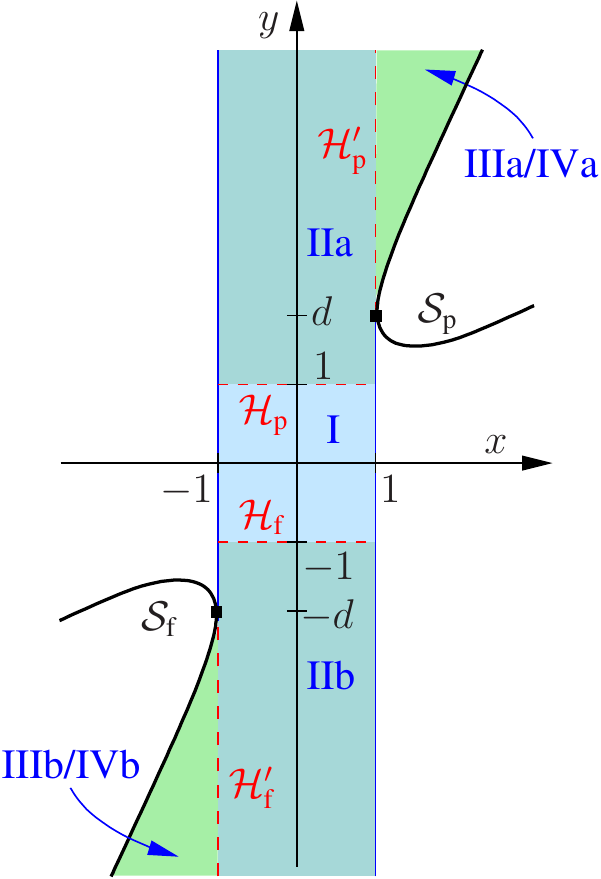}
 \caption{Extensions of the vacuum solution.\label{fig:extensions}}
\end{figure}

In the following we construct extensions of the solution into several regions which are shown in Fig.~\ref{fig:extensions}.

\subsubsection{Region I}
Our starting point is the Gowdy square, i.e.\ region I in Fig.~\ref{fig:extensions}. We rewrite the metric \eqref{eq:metric0} in terms of the coordinates $x=\cos\theta$, $y=\cos t$, $\rho_1$, $\rho_2$,
\begin{equation}
 \dd s^2=\ee^M\left(\frac{\dd x^2}{1-x^2}-\frac{\dd y^2}{1-y^2}\right)
         +R_0\left[(1-y^2)\ee^u\left(\dd\rho_1+Q\,\dd\rho_2\right)^2
              +(1-x^2)\ee^{-u}\dd\rho_2^2\right].
\end{equation}
The introduction of $y$ as a new time coordinate is particularly useful as this coordinate allows extensions through the past and future Cauchy horizons $\Hp$ and $\Hf$. As far as only these horizons are concerned, the extensions could be done in terms of the old coordinate $\theta$, 
but for later extensions into further regions it will be useful to also replace $\theta$ by $x$.

\subsubsection{Regions IIa and IIb}
First we consider the Cauchy horizon $\Hp$, which is generated by $\partial_{\rho_1}$ in our coordinates.
In order to remove the coordinate singularity at $y=1$, we replace the coordinate $\rho_1$ by $\rho_1'$ via
\begin{equation}
 \rho_1=\rho_1'+\kappa\ln|1-y|
\end{equation}
where the constant $\kappa$ has yet to be determined.
In terms of the coordinates $x$, $y$, $\rho_1'$ and $\rho_2$, the metric reads
\begin{eqnarray}
 \dd s^2 &=& \frac{R_0\kappa^2(1+y)^2\ee^u-\ee^M}{1-y^2}\,\dd y^2
             +\frac{\ee^M}{1-x^2}\,\dd x^2
             +R_0\Big[-2\kappa(1+y)\ee^u(\dd\rho_1'+Q\dd\rho_2)\dd y
  \nonumber\\
         &&  +(1-y^2)\ee^u(\dd\rho_1'+Q\,\dd\rho_2)^2+(1-x^2)\ee^{-u}\dd\rho_2^2\Big].
\end{eqnarray}
This metric is regular at $y=1$ and can be extended to $y>1$ (with analytical extensions of the potentials $u$, $M$, $Q$), provided we choose
\begin{equation}
 \kappa=\pm\sqrt{\lim\limits_{y\to1}\frac{\ee^{M-u}}{4R_0}}=\pm\frac{c_1}{2}.
\end{equation}
Hence we find two past extensions into region IIa in Fig.~\ref{fig:extensions}, corresponding to the two possible values of $\kappa$. (Note that, since $\kappa$ is independent of $d$ and $c_3$, it takes on the same values as for the past extensions of ``solutions 1 and 2'' discussed in \cite{BeyerHennig2014,Hennig2014}.) Once we have entered region IIa, we can use the inverse of the above coordinate transformation to recover the original form of the metric. This is then, of course, again singular at $y=1$ and therefore only valid for $y>1$. But since this form of the metric is simpler, it is preferable to use the original coordinates and accept that they cover only a certain part of the spacetime.

Similarly to the above construction, we can also regularize the metric at $y=-1$. This allows us to find future extensions through the future Cauchy horizon~$\Hf$, which is generated by $Q\f\partial_{\rho_1}-\partial_{\rho_2}$, where
\begin{equation}
 Q\f=\lim\limits_{y\to-1}Q=\frac{1 + c_3^2}{2 c_3}.
\end{equation}

To this end, we perform the coordinate transformation
\begin{equation}
 \rho_1=\rho_1'+\kappa_1\ln|1+y|,\quad
 \rho_2=\rho_2'+\kappa_2\ln|1+y|.
\end{equation}
Here, we find that the transformed metric is regular at $y=-1$ (and can therefore be extended into the region $y<-1$, i.e.\ region IIb in Fig.~\ref{fig:extensions}) if we choose the constants as
\begin{equation}
 \kappa_2=\pm\sqrt{\lim\limits_{y\to-1}\frac{(1-y^2)\ee^{M+u}}{4R_0(1-x^2)}}=\pm\frac{c_3 d^2}{c_1(d^2-1)}
\end{equation}
and
\begin{equation}
 \kappa_1=-\kappa_2\lim\limits_{y\to-1}Q=-\kappa_2 Q\f
         =\mp\frac{(1+c_3^2)d^2}{2c_1(d^2-1)}.
\end{equation}
Hence we again find two possible extensions, corresponding to the upper or lower choice of sign. Note that for $c_3=0$, the constant $\kappa_2$ vanishes so that the coordinate $\rho_2$ remains unchanged, $\rho_2'=\rho_2$. This is exactly what was observed for ``solution 2'' in \cite{Hennig2014}.

What can we say about the regularity of the solutions in regions IIa and IIb? Apart from the usual axes coordinate singularities\footnote{By way of example, it is shown in Appendix~\ref{sec:axiscoords} how the coordinate singularity at the axis part $x=1$, $1<y<d$ can be removed. To this end, suitable regular coordinates are introduced in a vicinity of this axis. Similar coordinate transformations can be done near the other axis parts.} at $x=\pm1$, a first possible source of irregularity is vanishing of the function $U$, which appears in the formulae for $\ee^u$ and $Q$, cf.~\eqref{eq:formulau}, \eqref{eq:formulaQ}. $U$ is always positive in the Gowdy square (region I), but it can indeed vanish along certain curves in regions IIa/b. As a consequence, $\ee^u\to0$, $\ee^{-u}\to\pm\infty$ and $Q\to\pm\infty$ there. However, in the particular combinations of these functions that appear in the metric coefficients [namely $Q\ee^u$ and $(1-y^2)\ee^u Q^2+(1-x^2)\ee^{-u}$] all inverse powers of $U$ cancel and these expressions are regular, as can easily be verified with the explicit solution. Hence the metric is regular at zeros of $U$ (provided they are not also zeros of $V$).

However, the solution is certainly expected to be singular at simultaneous zeros of $U$ and $V$, where $\ee^M$ becomes zero. This can be illustrated by computing the Kretschmann scalar near such a zero. To this end, using the exact solution, we observe that the metric has the structure
\begin{eqnarray}
 \dd s^2 &=& \alpha(U^2+V^2)\left(\frac{\dd x^2}{1-x^2}-\frac{\dd y^2}{1-y^2}\right)\\
 &&+\frac{\beta U}{U^2+V^2}\dd\rho_1^2+2\frac{\gamma U+\delta V^2}{U^2+V^2}\dd\rho_1\dd\rho_2
 + \frac{(\gamma U+\delta V^2)^2-A^2(U^2+V^2)^2}{\beta U(U^2+V^2)}\dd\rho_2^2,
\end{eqnarray}
where $\alpha$, $\beta$, $\gamma$, $\delta$, $A$ are functions of $x$ and $y$. The function $A$ must be given by $A=\sqrt{R_0^2(1-x^2)(y^2-1)}$ (which is real in the extensions) so that the $\rho_1$-$\rho_2$ part of the metric has the correct determinant. Moreover, since the component $g_{\rho_2\rho_2}$ is regular at $U=0$, $V\neq0$, as discussed above, the function $\delta$ must have the form $\delta=A+\eps U$ for some function $\eps=\eps(x,y)$. Expressing the Kretschmann scalar $K=R_{ijkl}R^{ijkl}$ in terms of these functions, we obtain
\begin{equation}\label{eq:Kret}
 K=\frac{P}{\alpha^6\beta^4A^4(U^2+V^2)^6},
\end{equation}
where $P$ is a lengthy polynomial in $x$, $y$ and $\alpha$, $\beta$, $\gamma$, $\eps$, $A$ and their $x$-$y$ derivatives up to second order.
If we finally replace $U$ and $V$ (but not their derivatives) in terms of polar coordinates\footnote{Note that the polar coordinates used here in the $U$-$V$ plane are not identical with the polar coordinates in the $x$-$y$ plane in Eq.~\eqref{eq:polar}. Apart from the fact that the former were introduced near the particular point  $x=1$, $y=-1$, whereas we consider general zeros of $U$ and $V$ here, the two descriptions differ by a nonlinear coordinate transformation $U=U(x,y)$, $V=V(x,y)$. However, near a simultaneous zero of $U$ and $V$, both radial coordinates are proportional to each other (with angular dependent proportionality constant) and hence provide comparable measures for the distance to the zero.}
by $U=r\cos\varphi$, $V=r\sin\varphi$, we observe $(U^2+V^2)^6=\mathcal O(r^{12})$, $P=r^4\tilde P$ with another polynomial $\tilde P$ and, therefore,
$K=\mathcal O(\tilde P r^{-8})$
near a simultaneous zero of $U$ and $V$ at $r=0$. The general form of the metric  in terms of $U$ and $V$ used here does not allow any  conclusions to be drawn about the behaviour of $\tilde P$ near $r=0$, but numerical studies show that $\tilde P$ itself vanishes at $r=0$ and behaves like $\mathcal O(r^2)$. This indicates that the Kretschmann scalar has the leading order behaviour
\begin{equation}\label{eq:Kret6}
 K=\mathcal O(r^{-6})
\end{equation}
near a singularity with $U=V=0$. 
Note that this observation is based on our numerical investigation of the function $\tilde P$ and hence not a rigorous result. It would be nice to find a proof to confirm \eqref{eq:Kret6} analytically, but due to the complicated structure of $\tilde P$ this may be rather difficult or even impossible.

For ``solution 1'' we found that there is always one zero of $\ee^M$ in the past extensions and there are either one or two zeros in the future extensions. For ``solution 2'', on the other hand, there are no zeros in the extensions. Therefore, we should expect that the present solution may or may not have singularities, depending on the choice of parameters. 

In a first step, we can easily verify that there cannot be zeros if we choose $c_3$ in the parameter region
\begin{equation}\label{eq:regpars}
 0\le c_3<\frac{1}{d+2}.
\end{equation}
The case $c_3=0$ leads to ``solution 2'', where we already know that there are no singularities. For $c_3\neq 0$, the condition $V=0$ implies that either $y=1$ or $r_d=d+2x-1/c_3$. The former case leads to a positive $U$ and can be excluded. The latter condition, together with \eqref{eq:regpars}, leads to $r_d=d+2x-1/c_3<d+2-(d+2)=0$. This is a contradiction since $r_d\ge 0$ must hold, thus establishing that $U=V=0$ has indeed no solution in the parameter region $\eqref{eq:regpars}$.

In a second step we consider the possibility of singularities outside this parameter region. Here we restrict ourselves to a numerical approach. The result of a thorough numerical discussion of zeros strongly indicates the following:
\begin{itemize}
 \item There is always one zero in the future extension if \eqref{eq:regpars} is violated.
 \item The parameter region without zeros for the past extension is larger than the region described by \eqref{eq:regpars}.
 \item Outside this larger regular parameter region, there are either one or two zeros in the past extension.
\end{itemize}

Hence we conclude that the solution is regular in regions IIa and IIb, with the possible exception of at most two curvature singularities at which $U=V=0$ holds.
(Note that the number of zeros was found with numerical investigations. If we restrict ourselves to analytical results, then we can only conclude that there are both parameter regions without and with singularities in the extensions, but in the latter case the exact number of singularities is unknown.)

\subsubsection{Regions IIIa and IIIb}

The boundaries $\Hp'$ ($x=1$, $y>d$) and $\Hf'$ ($x=-1$, $y<-d$) turn out to be further horizons though which the solution can be extended. 

First we consider $\Hp'$, which is generated by the Killing vector $Q\p'\partial_{\rho_1}-\partial_{\rho_2}$, where
\begin{equation}
 Q\p'=\lim\limits_{\begin{minipage}{8mm}
                   \scriptsize
                   $x\to1$\\ $y>d$
                   \end{minipage}}
                   Q=1+\frac{[1-c_2(2+d)]^2}{2c_3}.
\end{equation}
We can introduce coordinates in terms of which the solution is regular at $\Hp'$ as follows,
\begin{equation}
 \rho_1=\rho_1'+\mu_1\ln|1-x|,\quad
 \rho_2=\rho_2'+\mu_2\ln|1-x|.
\end{equation}
Similarly to the previous coordinate transformations, we can achieve that the transformed metric is regular at the horizon if we choose the constants appropriately. Here we obtain
\begin{equation}
 \mu_2= \pm\sqrt{\lim\limits_{x\to1}\left(-\frac{\ee^{M+u}}{4R_0}\right)}
      = \pm\frac{c_3 d^2}{2c_1(d+1)}
\end{equation}
and
\begin{equation}
 \mu_1=-Q\p'\mu_2=\mp\frac{d^2}{4c_1(d+1)}\left(2c_3+[1-c_3(2+d)]^2\right).
\end{equation}

Secondly, we extend the solution through $\Hf'$. This horizon is generated by $Q\f'\partial_{\rho_1}-\partial_{\rho_2}$, where
\begin{equation}
 Q\f'=\lim\limits_{\begin{minipage}{11mm}
                   \scriptsize
                   $x\to-1$\\ $y<-d$
                   \end{minipage}}
                   Q=-1+\frac{[1-c_2(d-2)]^2}{2c_3}.
\end{equation}

Regular coordinates can be obtained with the transformation
\begin{equation}
 \rho_1=\rho_1'+\nu_1\ln|1+x|,\quad
 \rho_2=\rho_2'+\nu_2\ln|1+x|.
\end{equation}
This time the requirement of regularity leads to
\begin{equation}
 \nu_2= \pm\sqrt{\lim\limits_{x\to-1}\left(-\frac{\ee^{M+u}}{4R_0}\right)}
      = \pm\frac{c_3 d^2}{2c_1(d-1)}
\end{equation}
and
\begin{equation}
 \nu_1=-Q\f'\nu_2=\pm\frac{d^2}{4c_1(d-1)}\left(2c_3-[1-c_3(d-2)]^2\right).
\end{equation}

The solution could potentially be singular in regions IIIa/b if $U$ and $V$ had simultaneous zeros there. This, however, is not the case, which can be seen as follows. In these regions we have $1-x^2<0$ and $1-y^2<0$, so that
$r_d^2=(d-xy)^2-(1-x^2)(1-y^2)<(d-xy)^2$ and $r_d<xy-d$.
Consequently, $d+r_d-xy<0$ and $d-r_d-xy<-2r_d<0$. Using these estimates, we see that the square bracket in the formula \eqref{eq:defU2} for $U$ is negative, i.e.\ $U$ has no zeros. 

The only potential irregularities come from zeros of $r_d$ (at which $\ee^M$ diverges). Indeed, the equation $r_d^2=0$ defines a hyperbola in the $x$-$y$ plane with the two branches $\Sp$ and $\Sf$, cf.\ Fig.~\ref{fig:extensions}. But similarly to the discussion of ``solution 2'' in \cite{Hennig2014}, we will see that we only have coordinate singularities there. This issue is described in the next subsection. 

\subsubsection{Regions IVa and IVb}
Finally, we remove the coordinate singularity at the hyperbola $\Sp$. (A similar transformation can be used to obtain regular coordinates in a neighbourhood of $\Sf$.) To this end, we introduce null coordinates $\tilde\alpha$ and $\tilde\beta$ via
\begin{eqnarray}
 x&=&\sqrt{\frac12\left(\alpha\beta+1-\sqrt{(\alpha^2-1)(\beta^2-1)}\right)},\\
 y&=&\sqrt{\frac12\left(\alpha\beta+1+\sqrt{(\alpha^2-1)(\beta^2-1)}\right)}
\end{eqnarray}
and
\begin{equation}
 \tilde\alpha=\sqrt{\alpha-d},\quad
 \tilde\beta=\sqrt{\beta-d}.
\end{equation}
The upper half of $\Sp$ is given by $\tilde\alpha=0$, $\tilde\beta>0$, and the region above the hyperbola corresponds to positive values of $\tilde\alpha$. In these coordinates, the formula for $r_d$ simplifies to $r_d=\tilde\alpha\tilde\beta$, and the originally singular part of the metric becomes
\begin{equation}
 \ee^M\left(\frac{\dd x^2}{1-x^2}-\frac{\dd y^2}{1-y^2}\right)
 \propto\frac{1}{r_d}\left(\frac{\dd x^2}{1-x^2}-\frac{\dd y^2}{1-y^2}\right)
 \propto\dd\tilde\alpha\,\dd\tilde\beta.
\end{equation}
It is now possible to extend the metric through the singularity by allowing $\tilde\alpha\le0$. This corresponds to region IVa in Fig.~\ref{fig:extensions}.

As before, afterwards we can transform back to our earlier coordinates $x$ and $y$, in order to retrieve the simpler form of the metric. Note that region IVa corresponds to the same coordinate domain in the $x$-$y$ plane as region IIIa. (Similarly, region IVb ``beyond $\Sf$'' corresponds to the same coordinate domain as region IIIb.) Nevertheless, the metric is still different in these regions as the square root $r_d$ has to be taken with the opposite sign in the new regions. It also follows that $x$ and $y$ swap their roles as time and space coordinates. A particular consequence is that the hyperbola $\Sf$ is in the past of any observer in region IVb, so that it is not possible to return to $\Sf$.

Similarly to the argument in the previous section, we can again easily check that there are no singularities in these regions. 

Hence we have found that the causal structure of the present solution is the same as that of ``solution 2''. However, while ``solution 2'' has no singularities, the extensions here can contain singularities in regions IIa and IIb.

\section{A new family of electrovacuum solutions}\label{sec:newelectrovac}
Now we leave the realm of vacuum solutions and intend to obtain interesting examples of cosmological models with a non-trivial additional field. In particular, we  construct a family of exact solutions to the Gowdy-symmetric Einstein-Maxwell equations. To this end, we first formulate the relevant equations and then describe the solution process.

\subsection{Metric and field equations}
In the presence of an electromagnetic field, we can still use the line element from our previous discussions of the vacuum case. We choose coordinates $(x^i)=(x,y,\rho_1,\rho_2)$, i.e.\ we immediately replace $t$ and $\theta$ by $x=\cos\theta$ and $y=\cos t$, since these coordinates are more convenient for the construction of extensions through the Cauchy horizons. Then the line element reads
  \begin{equation}\label{eq:metric}
   \dd s^2 =\ee^M\left(\frac{\dd x^2}{1-x^2}-\frac{\dd y^2}{1-y^2}\right)+R_0\left[(1-y^2)\,\ee^u 
   (\dd\rho_1+Q\,\dd\rho_2)^2+(1-x^2)\,\ee^{-u} \dd\rho_2^2\right].
  \end{equation}
The electromagnetic field is described in terms of the field tensor
\begin{equation}
 F_{ij}=A_{i,j}-A_{j,i}, 
\end{equation}
where we choose an electromagnetic 4-potential $A_i$ of the form
\begin{equation}
    (A_i)=(0,0,A_3,A_4),
\end{equation}
which is compatible with our requirement of Gowdy-symmetry.
The corresponding field tensor is 
\begin{equation}
    (F_{ij})=\left(\begin{array}{cccc}
              0       & 0       & -A_{3,x} & -A_{4,x}\\
              0       & 0       & -A_{3,y} & -A_{4,y}\\
              A_{3,x} & A_{3,y} & 0        & 0\\
              A_{4,x} & A_{4,y} & 0        & 0
             \end{array}\right).
\end{equation}

We then have to solve Einstein's field equations 
with the energy-momentum tensor
\begin{equation}\label{eq:Tij}
 T_{ij}=\frac{1}{4\pi}(F_{ki}F^k{}_j-\frac14 g_{ij}F_{kl}F^{kl}).
\end{equation}
Since this tensor satisfies $T=T^i{}_i=0$, the field equations are
$R_{ij}=8\pi T_{ij}$. The resulting equations are listed in Appendix~\ref{sec:EM}, cf.~\eqref{eq:ueq}--\eqref{eq:My}.

Finally, we also have to satisfy Maxwell's equations $\nabla_{j}F^{ij}=0$. The explicit equations can again be found in Appendix~\ref{sec:EM}.

\subsection{Ernst formulation}
Similarly to the Gowdy-symmetric Einstein vacuum equations, whose essential part can be reformulated in terms of the single, complex Ernst equation \eqref{eq:EE}, the essential part of the combined Einstein-Maxwell equations can be reformulated in terms of two complex Ernst equations for two Ernst potentials $\Phi$ and $\E$.

For that purpose, we define quantities $f$ and $a$ in terms of the two Killing vectors via
\begin{eqnarray}
  \label{eq:f}
  f &=& \frac{1}{R_0}g(\partial_{\rho_2},\partial_{\rho_2})
        = Q^2\ee^u(1-y^2)+\ee^{-u}(1-x^2),\\
  \label{eq:a}
  a &=& \frac{g(\partial_{\rho_1},\partial_{\rho_2})}
             {g(\partial_{\rho_2},\partial_{\rho_2})}
        = \frac{Q}{f}\ee^u(1-y^2).
\end{eqnarray}
We also define a quantity $\beta$ in terms of $f$ and the 4-potential $A_i$ as a solution to the first-order equations
\begin{eqnarray}
 \label{eq:betax}
 \beta_{,x} &=& \frac{f}{1-x^2}(A_{3,y}-aA_{4,y}),\\
 \label{eq:betay}
 \beta_{,y} &=& \frac{f}{1-y^2}(A_{3,x}-aA_{4,x}).  
 \end{eqnarray}
(Here we can make an arbitrary choice of the integration constant, since only derivatives of $\beta$ are physically relevant.)
From the fourth component (the $\rho_2$-component) $A_4$ of the 4-potential and this function $\beta$, we construct the first Ernst potential
\begin{equation}\label{eq:Phi}
 \Phi = \frac{1}{\sqrt{R_0}}(A_4+\ii\beta).
\end{equation}
Furthermore, we define a quantity $b$ as a solution to
\begin{eqnarray}
  \label{eq:ax}
  a_{,x} &=& \frac{1-y^2}{f^2}\left[b_{,y}
             +\ii(\bar\Phi\Phi_{,y}-\Phi\bar\Phi_{,y})\right],\\
  \label{eq:ay}
  a_{,y} &=& \frac{1-x^2}{f^2}\left[b_{,x}
             +\ii(\bar\Phi\Phi_{,x}-\Phi\bar\Phi_{,x})\right],
\end{eqnarray}
where the integration constant is again irrelevant. This allows us to define the second Ernst potential 
\begin{equation}\label{eq:E}
  \E   = f+|\Phi|^2+\ii b.
\end{equation}

As a consequence of the Einstein-Maxwell equations, $\Phi$ and $\E$ satisfy the Ernst equations
\begin{eqnarray}
   \nonumber
   &&f\cdot\left[(1-x^2)\E_{,xx}-2x\E_{,x}-(1-y^2)\E_{,yy}+2y\E_{,y}\right]\\
   &&   \qquad = (1-x^2)(\E_{,x}-2\bar\Phi\Phi_{,x})\E_{,x}
       -(1-y^2)(\E_{,y}-2\bar\Phi\Phi_{,y})\E_{,y},\label{eq:Ernst1}\\
   \nonumber
   &&f\cdot\left[(1-x^2)\Phi_{,xx}-2x\Phi_{,x}
          -(1-y^2)\Phi_{,yy}+2y\Phi_{,y}\right]\\
   &&   \qquad = (1-x^2)(\E_{,x}-2\bar\Phi\Phi_{,x})\Phi_{,x}
       -(1-y^2)(\E_{,y}-2\bar\Phi\Phi_{,y})\Phi_{,y}.\label{eq:Ernst2}
\end{eqnarray}
Conversely, if $\Phi$ and $\E$ are solutions to the Ernst equations, then we can construct the corresponding metric and electromagnetic potentials. First we obtain $A_4$ and $\beta$ from \eqref{eq:Phi}, and $f$ and $b$ from \eqref{eq:E}. The integrability condition of \eqref{eq:ax}, \eqref{eq:ay} is satisfied as a consequence of the Ernst equations, which allows us to solve these equations for $a$. From $f$ and $a$ we can construct the metric potentials $u$ and $Q$ using \eqref{eq:f}, \eqref{eq:a}. Then we solve \eqref{eq:betax}, \eqref{eq:betay} for $A_3$ and \eqref{eq:Mx}, \eqref{eq:My} for $M$. The integrability conditions of these equations are again satisfied as a consequence of the Ernst equations. Note that the integration constant in the computation of $A_3$ is irrelevant since the 4-potential is only defined up to an additive constant, and the integration constant for the $M$-equations can be fixed by the conditions for regularity on the axis.

\subsection{Construction of a family of exact solutions}
In order to solve the electrovacuum Ernst equations \eqref{eq:Ernst1}, \eqref{eq:Ernst2}, we could choose some initial potentials, i.e.\ specify $\E$ and $\Phi$ at $t=0$ ($y=1$). Then we would need to solve an initial value problem, which could again be done with Sibgatullin's integral method. (This method is not restricted to the case of vacuum, but covers the electrovacuum case as well.) However, since we have already constructed some vacuum solutions, there is a quicker and simpler way to obtain an electrovacuum solution: we can apply a Harrison transformation \cite{Harrison1968}, which maps vacuum into electrovacuum solutions. To this end, we start from an arbitrary solution $\E$ to the vacuum Ernst equation. Then $\E'$ and $\Phi'$, defined by\footnote{Our formulae for the Harrison transformation have plus signs in the denominators instead of the minus sign in the form of the Harrison transformation as originally applied to axisymmetric and stationary spacetimes. This is related to our definitions of the Ernst potentials for Gowdy spacetimes with two spacelike Killing vectors, as opposed to the situation with one timelike and one spacelike Killing vector in axisymmetric and stationary spacetimes.}
\begin{equation}\label{eq:HT}
  \E'=\frac{\E}{1+|\gamma|^2\E},\quad
  \Phi'=\frac{\gamma\E}{1+|\gamma|^2\E},\quad
  \gamma\in\C,
\end{equation}
satisfy the electrovacuum Ernst equations \eqref{eq:Ernst1}, \eqref{eq:Ernst2}.
Note that for $\gamma=0$ this becomes the identity transformation $\E'=\E$, $\Phi'=0$, but for $\gamma\neq 0$ we obtain non-trivial new potentials, corresponding to new solutions with electromagnetic field ($\Phi'\neq0$).

We apply this transformation to ``solution 1''.\footnote{We could apply the same transformation to ``solution 2'', or even to the combined family derived above, but this is expected to be technically very challenging. Hence we restrict ourselves to the probably simplest case of ``solution 1'', where the Ernst potential is a rational function in $x$ and $y$ and does not involve the square root $r_d$.}
Moreover, we can restrict ourselves\footnote{For complex $\gamma$ of the form $\gamma=g\ee^{\ii\phi}$, the transformation \eqref{eq:HT} becomes $\E\mapsto\frac{\E}{1+g^2\E}$, $\Phi\mapsto \frac{g\ee^{\ii\phi}\Phi}{1+g^2\E^2}$. This can be decomposed into the two transformations $\E\mapsto\frac{\E}{1+g^2\E}$, $\Phi\mapsto \frac{g\Phi}{1+g^2\E^2}$ and $\E\mapsto\E$, $\Phi\mapsto\ee^{\ii\phi}\Phi$. The first of these is a Harrison transformation with \emph{real} parameter $\gamma=g$, and the second is a special case of the transformation $\E\mapsto \alpha\bar\alpha\E$, $\Phi\mapsto\alpha\Phi$, which merely corresponds to a constant  duality rotation of the Maxwell field 
and leaves the energy-momentum tensor invariant, see \cite{Stephani, Witten}. This justifies considering only the first of these two transformations.}
to real values of $\gamma$. Then we obtain
\begin{equation}
  \E'=\frac{\E}{1+\gamma^2\E},\quad
  \Phi'=\frac{\gamma\E}{1+\gamma^2\E},\quad
  \gamma\in\R,
\end{equation}
where $\E$ is the Ernst potential of ``solution 1'', which can be found in \cite{BeyerHennig2014}, cf.~Eq.~(39) there.

In a next step we construct the corresponding metric and electromagnetic potentials. The required calculations are very lengthy, but eventually all integrals can be computed and exact formulae for all functions are obtained. Unfortunately, in contrast to the previously discussed solutions, in this computation not only the intermediate results but also the final functions are very lengthy expressions. (At least I was not able to identify suitable abbreviations for terms that appear repeatedly, such that the formulae become concise expressions in terms of these abbreviations.) Therefore, we do not give the full expressions here\footnote{Please email the author if you would like to obtain a Mathematica notebook with the full solution.}, but restrict ourselves to the lowest-order corrections in $\gamma$ to the potentials for ``solution 1''. The result is
\begin{eqnarray*}
  \ee^u &=& \frac{16c_1 U}{(1 + y)(U^2 + V^2)}
           \!\left[1+2\gamma^2\left((1-x^2)(1+y)\frac{U^2 + V^2}{16c_1U} 
                -(1-y)\frac{16c_1 U}{U^2+V^2}\tilde Q^2\right)\!\right] 
             \!\!+\mathcal O(\gamma^4),\\
  Q &=& \left[1 - \gamma^2(1-x^2)(1+y)\frac{U^2+V^2}{4c_1 U}\right]\tilde Q
         +\mathcal O(\gamma^4),\\
  \ee^M &=&  \frac{R_0(U^2 + V^2)}{64c_1^3}
     \left[1 + 2 \gamma^2\left((1-x^2)(1+y)\frac{U^2+V^2}{16c_1U}
                + (1-y)\frac{16c_1U}{U^2+V^2}\tilde Q^2\right)\right]
             +\mathcal O(\gamma^4),\\
  A_3 &=& \gamma\sqrt{R_0} (1-y)\frac{16c_1U}{U^2+V^2}\tilde Q 
         +\mathcal O(\gamma^3),\\
  A_4 &=& \gamma\sqrt{R_0}\left[(1-x^2)(1+y)\frac{U^2+V^2}{16c_1U}
             + (1-y)\frac{16c_1U}{U^2 + V^2}\tilde Q^2\right]
              +\mathcal O(\gamma^3),
 \end{eqnarray*}
 where
 \begin{equation}
  U = 4c_1^2(1+y) + c_3^2(1-x^2)(1-y)^3,\quad
  V = 4c_1(1-y)[1-c_3 x(2+y)],
 \end{equation}
 \begin{equation}
  \tilde Q = x + \frac{c_3}{8}(1-x^2)\left(7+4y+y^2 
    + \frac{(1-y)V^2}{4c_1^2U}\right).
 \end{equation}
The integration constant in the $M$-integration has been chosen so that $\ee^{u+M}\big|_{x=\pm1}=R_0+\mathcal O(\gamma^4)$. Hence, to the discussed order in $\gamma$, the solution satisfies the axes boundary conditions of solutions with a past Cauchy horizon with \emph{closed} null generators, see \eqref{eq:regcond1}. Moreover, the (irrelevant) integration constant in the $A_3$-integration has been chosen so that $A_3|_{y=1}=0$.

\subsection{Properties of the solution}

An investigation of the solution near $y=\pm1$ reveals that these boundaries correspond to Cauchy horizons through which the solution can be extended. The transformation to regular coordinates, in which the extensions can be done, is given below. This shows, in particular, that the Cauchy horizons $\Hp$ and $\Hf$ of ``solution 1'' (which is contained in this solution family for $\gamma=0$) are not ``destroyed'' when we modify the solution by including an electromagnetic field (which is present for $\gamma\neq 0$).

Since our solution contains a non-trivial additional field, namely an electromagnetic field, it is interesting to look at the energy density $\eps$ of that field. From the point of view of an observer with fixed spatial coordinates $x$, $\rho_1$, $\rho_2$, i.e.\ an observer with 4-velocity 
$(u^i)=(0,\sqrt{1-y^2}\,\ee^{-M/2},0,0)$, the energy density of the electromagnetic field is $\eps= T_{ij}u^iu^j=(1-y^2)\ee^{-M}T_{yy}$. With the energy-momentum tensor \eqref{eq:Tij} and our form of the field tensor, this becomes
\begin{equation}
 \eps=\frac{ \ee^{-M-u}}{8\pi R_0} 
    \left[\frac{1-x^2}{1-y^2}(A_{3,x})^{2} +(A_{3,y})^{2}
   +\ee^{2u}\left((A_{4,x}-Q A_{3,x})^2+\frac{1-y^2}{1-x^2}(A_{4,y}-Q A_{3,y})^2\right)\right].
\end{equation}
We can further specialize this formula to our solution by plugging in the expressions for the metric and electromagnetic potentials from the previous subsection. Here we give only the structure of the result, which turns out to be of the form
\begin{equation}
 \eps=\gamma^2\frac{p_{8,10}}{R_0(U^2+V^2)^2}+\mathcal O(\gamma^4),
\end{equation}
where $p_{8,10}$ is a polynomial of 8th degree in $x$ and 10th degree in $y$. (In adition, $p_{8,10}$ also depends on the parameters $c_1$ and $c_3$, but it is independent of $\gamma$ and $R_0$.)

Obviously, the energy density is bounded except at simultaneous zeros of $U$ and $V$. As we will see below, such a zero is equivalent to a diverging Kretschmann scalar (exactly as in the above discussion of the exact \emph{vacuum} solution). Hence the energy density diverges at curvature singularities of the solution, which are present in the ``singular cases'' and in the extended regions beyond the Cauchy horizons. 

The Kretschmann scalar for this solution has the following form,
   $$K=R_{ijkl}R^{ijkl}=\frac{p_{16,24}}{R_0^2(U^2+V^2)^6}
      +\gamma^2\frac{p_{24,34}}{R_0^2(U^2+V^2)^7}+\mathcal O(\gamma^4),$$
where $p_{n,m}$ again denotes polynomials of degree $n$ in $x$ and degree $m$ in $y$.
We see that, similarly to the energy density, the Kretschmann scalar is everywhere regular where $U$ and $V$ do not vanish simultaneously --- exactly as for ``solution 1''. 

Another measure of regularity are the Ernst potentials themselves, since they are invariantly defined in terms of the Killing vectors. Due to
\begin{equation}
 \E'=(1-\gamma^2\E)\E+\mathcal O(\gamma^4),\quad
 \Phi'=\gamma\E'
\end{equation}
and since $\E$ turns out to have the form
\begin{equation}
 \E=\frac{p_{6,7}}{U-\ii V},
\end{equation}
the regularity of $\E'$ and $\Phi'$, to the order considered here, is also guaranteed wherever $U$ and $V$ do not vanish simultaneously.

The electromagnetic solution turns out to have the same causal structure as ``solution 1''. In particular, we can extend the solution through the past and future Cauchy horizons, but beyond these there are no further horizons. In order to construct these extensions, we can use the following coordinate transformations:
 \begin{itemize}
  \item Extension through $\Hp$:
   $$\rho_1=\rho_1'+\kappa\ln|1-y|,\quad
     \kappa=\pm\frac{c_1}{2}.$$
  \item Extension through $\Hf$:
   $$\rho_1=\rho_1'+\kappa_1\ln|1+y|,\quad
     \rho_2=\rho_2'+\kappa_2\ln|1+y|,\quad
     \kappa_2=\pm\frac{c_3}{c_1},\quad
     \kappa_1=\mp\frac{1+c_3^2}{2c_1}.$$
 \end{itemize}
 Note that, to the order considered here, $\gamma$ does not explicitly appear in these transformations. Nevertheless, the presence of Cauchy horizons and the constructions of extensions is not restricted to the expansion in terms of $\gamma$. If we repeat the above calculations for the \emph{full} metric, we see that we can still use the above coordinate transformations to construct extensions, provided we replace the constants $\kappa$, $\kappa_1$, $\kappa_2$ with
 \begin{eqnarray}
  \kappa &=& \pm\frac{c_1 [1 + (1 + 4 c_3^2) \gamma^4]}{2 [1 + 2(1 + 4 c_3^2) \gamma^4 + (1 -  4 c_3^2)^2 \gamma^8]},\\
  \kappa_2 &=& \pm\frac{c_3 [1 + (5 - 4 c_3^2) \gamma^4] [1 + (1 + 
      4 c_3^2) \gamma^4]}{c_1 [1 + 2(1 + 4 c_3^2) \gamma^4 + (1 - 
      4 c_3^2)^2 \gamma^8]},\\
  \kappa_1 &=& \mp \frac{(1 + c_3^2) [1 + (1 + 4 c_3^2) \gamma^4]}
   {2 c_1 [1 + 2(1 + 4 c_3^2) \gamma^4 + (1 - 4 c_3^2)^2 \gamma^8]}.
 \end{eqnarray}
This shows how $\gamma$ enters these transformations in higher orders.

\section{Discussion}\label{sec:discussion}

We have constructed a new family of explicit inhomogeneous cosmological vacuum models within the class of smooth Gowdy-symmetric generalised Taub-NUT (SGGTN) solutions. The exact solution depends on the four parameters $R_0>0$, $c_1>0$, $c_3\in\R$ and $d>1$. In the limit where both $d\to\infty$ and $c_3\to 0$, the solution reduces to the well-known Taub-NUT solution. If we perform the individual limit $c_3\to0$, this leads to the 3-parameter models derived in \cite{BeyerHennig2014} (denoted as ``solution 1'' above), whereas the limit $d\to\infty$ produces the 3-parameter models introduced in \cite{Hennig2014} (``solution 2''). Hence we have been able to embed two previously known solutions into a larger family of solutions. Our new solution is generally (for $c_3\neq\pm1$) regular in the maximal globally hyperbolic region, i.e.\ in the ``Gowdy square''. In the ``singular cases'' $c_3=\pm1$ on the other hand, curvature singularities form at the future boundary of the Gowdy square. These exhibit the same directional behaviour that had been observed for the singularities in ``solution 1''. The future and past boundaries of the Gowdy square are Cauchy horizons through which the solution can be extended. Beyond these Cauchy horizons, we have found further horizons, and again the solution can be extended. All these extensions have been explicitly constructed, and we have shown that, depending on the parameter values, they may or may not contain curvature singularities. If singularities are present, then our numerical investigations indicate that there are either one or two points where the Kretschmann scalar diverges.

Furthermore, we have derived the equations that describe a more general class of SGGTN solutions, namely Gowdy-symmetric solutions with an additional electromagnetic field. By applying a Harrison transformation to ``solution 1'', we have obtained an exact 4-parameter family of solutions for this case. These solutions still have Cauchy horizons at the boundaries of the Gowdy square, i.e.\ we observe that the horizons in ``solution 1'' do not disappear if we distort this solution with an electromagnetic field. Moreover, we have studied the singularity structure and the causal structure for this solution family and found these to be the same as for ``solution 1''.

For future investigations of properties of cosmological models and for probing strong cosmic censorship, these families of exact solutions may prove very useful, as they could be used as background models for perturbation analyses. It would be interesting to investigate numerically how these exact solutions behave under perturbations that lead to deviations from the ideal Gowdy symmetry, or how they behave if matter fields are included. If we believe in strong cosmic censorship, we should expect that the Cauchy horizons and the acausal regions behind them are unstable and disappear in these modified models.

\section*{Acknowledgments}
 I would like to thank Gerrard Liddell for commenting on the manuscript. This work was supported by the Marsden Fund Council from Government funding, administered by the Royal Society of New Zealand.
\appendix
\section{Regular coordinates near the axis\label{sec:axiscoords}}

We intend to construct regular coordinates near the part $x=1$, $1<y<d$ of the axis for the vacuum solution derived in Sec.~\ref{sec:vacuum}. To this end, we first change the Killing basis back to $\partial_{\tilde\rho_1}$, $\partial_{\tilde\rho_2}$, where $\tilde\rho_1$ and $\tilde\rho_2$ are the coordinates that we originally used for the construction of smooth Gowdy-symmetric generalised Taub-NUT solutions in \cite{BeyerHennig2012}.\footnote{These were denoted as $\partial_{\rho_1}$ and $\partial_{\rho_2}$ without tildes in \cite{BeyerHennig2012}.} These are related to our current coordinates $\rho_1$, $\rho_2$ via
\begin{equation}
 \tilde\rho_1 = \rho_1-a\p\rho_2,\quad
 \tilde\rho_2 = -a\p\rho_1+\rho_2, 
 \quad\textrm{where}\quad a\p=-\frac{1}{d},
\end{equation}
cf.~Eqs.~(32), (36), (42) in \cite{Hennig2014}.
The inverse transformation is of the form 
\begin{equation}
 \rho_1=\alpha\tilde\rho_1+\beta\tilde\rho_2,\quad
 \rho_2=\gamma\tilde\rho_1+\delta\tilde\rho_2
\end{equation}
with
\begin{equation}
 \alpha=\delta=\frac{1}{1-a\p^2},\quad
 \beta=\gamma=\frac{a\p}{1-a\p^2}.
\end{equation}
The metric in terms of $\tilde\rho_1$, $\tilde\rho_2$, which are defined in the domain
\begin{equation}\label{eq:domain}
 \frac{\tilde\rho_1+\tilde\rho_2}{2}\in(0,2\pi),\quad
 \frac{\tilde\rho_1-\tilde\rho_2}{2}\in(0,2\pi),
\end{equation}
has the same form as before, but with new potentials $\tilde u$, $\tilde Q$ and a new constant $\tilde R_0$. Applying the general formulae for rotations of the Killing basis in \cite{BeyerHennig2012}, we find
\begin{eqnarray}
 \tilde R_0 &=& |\alpha\delta-\beta\gamma|R_0=\frac{R_0}{1-a\p^2},\\
 \ee^{\tilde u} &=& 
 \frac{(1-y^2)(\alpha+\gamma Q)^2\ee^u+\gamma^2(1-x^2)\ee^{-u}}
 {|\alpha\delta-\beta\gamma|(1-y^2)}\nonumber\\
 &=& \frac{(1-y^2)(1+a\p Q)^2\ee^u+a\p^2(1-x^2)\ee^{-u}}{(1-a\p^2)(1-y^2)},\\
 \tilde Q &=& \frac{(1-y^2)(\alpha+\gamma Q)(\beta+\delta Q)+\gamma\delta(1-x^2)\ee^{-2u}}{(1-y^2)(\alpha+\gamma Q)^2+\gamma^2(1-x^2)\ee^{-2u}}\nonumber\\
 &=& \frac{(1-y^2)(1+a\p Q)(a\p+Q)+a\p(1-x^2)\ee^{-2u}}{(1-y^2)(1+a\p Q)^2+a\p^2(1-x^2)\ee^{-2u}}.
\end{eqnarray}
The boundary value $Q=1$ on the axis under consideration is unchanged by this transformation, i.e.\ we also have $\tilde Q=1$ on this axis. Moreover, the boundary condition $\ee^{M+\tilde u}=\tilde R_0$ holds on the axis. Using this condition, it follows that the additional coordinate transformation
\begin{equation}\label{eq:axistransform}
 p=\sqrt{1-x^2}\cos(\tilde\rho_2),\quad
 q=\sqrt{1-x^2}\sin(\tilde\rho_2),\quad
 \tau=\tilde\rho_1+\tilde\rho_2
\end{equation}
leads to
\begin{equation}
 \dd s^2=\ee^M\left(\dd p^2+\dd q^2+\frac{\dd y^2}{1-y^2}\right)
         -\tilde R_0(y^2-1)\ee^{\tilde u}\dd\tau^2+\mathcal O(1-x), 
\end{equation}
i.e.\ the metric in terms of $p$, $q$, $y$, $\tau$ is regular in a neighbourhood of the axis.

\begin{figure}\centering
 \includegraphics[scale=0.7]{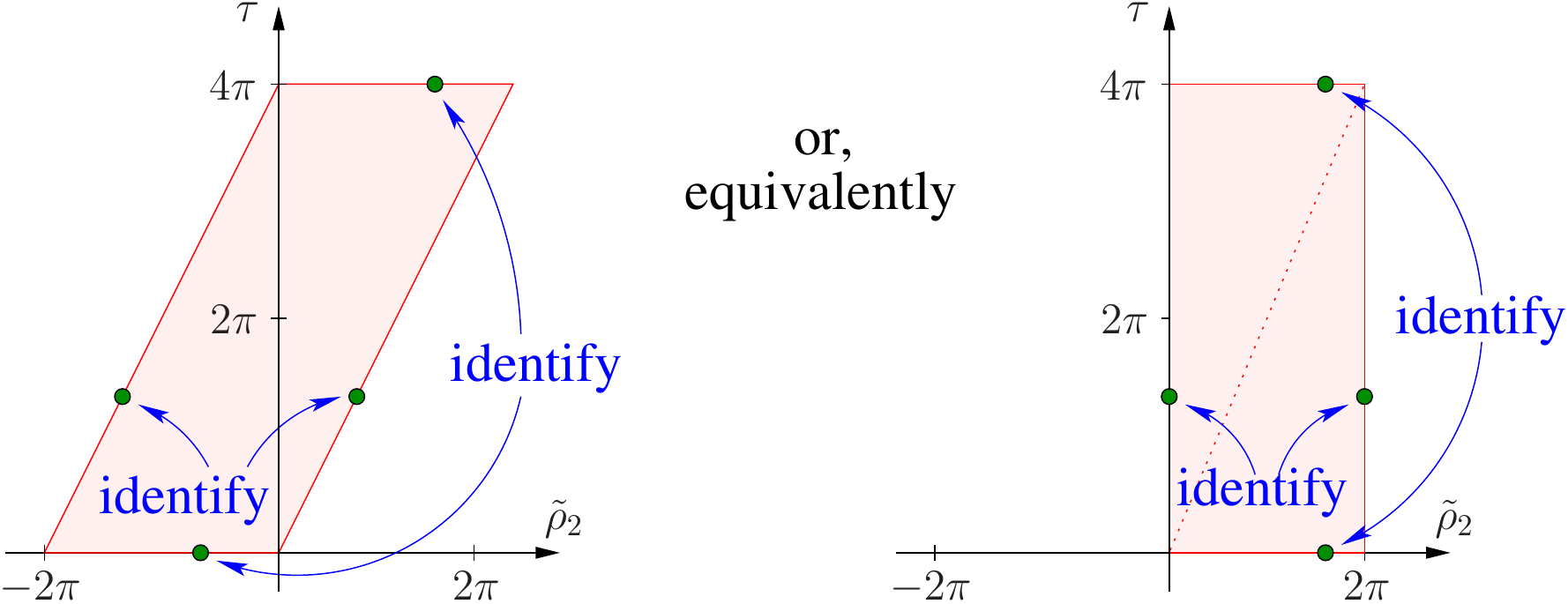}
 \caption{Domain for the coordinates $\tilde\rho_2$ and $\tau$.\label{fig:domain}}
\end{figure}

Note that the acausal character of the solution in this extension (existence of closed causal curves) shows up in the periodic nature of the new time coordinate $\tau$, which can be seen as follows. Due to \eqref{eq:domain}, the coordinates $\tilde\rho_2$ and $\tau$ are defined in the parallelogram-shaped region shown in Fig.~\ref{fig:domain}, left panel. 
Because of the periodic nature of the coordinates, points on opposite sides of the parallelogram have to be identified as indicated in the figure. As a consequence, we can rearrange the domain to obtain the rectangular region shown in the right panel (by shifting points in the left triangular half of the parallelogram by $2\pi$ to the right). Hence we have achieved that $\tilde\rho_2\in(0,2\pi)$ [which justifies that $\tilde\rho_2$ can be used as an angular coordinate in the transformation \eqref{eq:axistransform}] and $\tau\in(0,4\pi)$. Due to the $4\pi$-periodicity of $\tau$, a curve with fixed spatial coordinates $p$, $q$, $y$ and variable time coordinate $\tau$ is hence seen to be a closed timelike curve.

\section{Einstein-Maxwell equations}\label{sec:EM}
Here we list the field equations for Gowdy-symmetric electrovacuum solutions.
In terms of the coordinates $x$, $y$ and the metric potentials appearing in the line element \eqref{eq:metric}, Einstein's field equations can be reduced to a second-order equation for $u$,
\begin{eqnarray}
     \nonumber
     &&(1-x^2)u_{,xx}-(1-y^2)u_{,yy} -\frac{1-y^2}{1-x^2}\,\ee^{2u}
      \left[(1-x^2)Q_{,x}^{\ 2}-(1-y^2)Q_{,y}^{\ 2}\right]
       -2xu_{,x}+2yu_{,y}+2\\
     && \nonumber
      \quad = \frac{2}{R_0}\left(\frac{\ee^u}{1-x^2}
       \left[(1-x^2)(A_{4,x}-QA_{3,x})^2-(1-y^2)(A_{4,y}-QA_{3,y})^2\right]
       \right.\\
     &&  \qquad\left.
       -\frac{\ee^{-u}}{1-y^2}\left[(1-x^2)A_{3,x}^{\ 2}-(1-y^2)A_{3,y}^{\ 2}\right]\right),\label{eq:ueq}
    \end{eqnarray}
a second-order equation for $Q$,
\begin{eqnarray}
    \nonumber
    &&(1-x^2)Q_{,xx}-(1-y^2)Q_{,yy}+2(1-x^2)Q_{,x}u_{,x}
    -2(1-y^2)Q_{,y}u_{,y}+4yQ_{,y}\\
    &&\quad  =-\frac{4\ee^{-u}}{R_0(1-y^2)}
    \left[(1-x^2)(A_{4,x}-QA_{3,x})A_{3,x}-(1-y^2)(A_{4,y}-QA_{3,y})A_{3,y}\right],\label{eq:Qeq}
   \end{eqnarray}
and two first-order equations for $M$,
\begin{eqnarray}\label{eq:Mx}
   M_{,x} &=& -\frac{1-y^2}{2(x^2-y^2)}\Bigg[
    x(1-x^2)u_{,x}^{\ 2}+x(1-y^2)u_{,y}^{\ 2}-2y(1-x^2)u_{,x}u_{,y}
     \nonumber\\
   &&
   +2\frac{x^2+y^2-2x^2y^2}{1-y^2}u_{,x}-4xyu_{,y}-4x\nonumber\\
   &&
   +\frac{1-y^2}{1-x^2}\ee^{2u}
   \Big(x(1-x^2)Q_{,x}^{\ 2}+ x(1-y^2)Q_{,y}^{\ 2}
   -2y(1-x^2)Q_{,x}Q_{,y}\Big)\nonumber\\  
   &&
   +\frac{4\ee^u}{(1-x^2)R_0}\Big[\frac{1-x^2}{1-y^2}\ee^{-2u}
   \Big(x(1-x^2)A_{3,x}^{\ 2}+x(1-y^2)A_{3,y}^{\ 2}
    -2y(1-x^2)A_{3,x}A_{3,y}\Big)\nonumber\\
   &&
   +x(1-x^2)(A_{4,x}-QA_{3,x})^2+x(1-y^2)(A_{4,y}-QA_{3,y})^2\nonumber\\
   &&
   -2y(1-x^2)(A_{4,x}-QA_{3,x})(A_{4,y}-QA_{3,y})\Big]\Bigg],\label{eq:Meq1}
  \end{eqnarray}
 \begin{eqnarray}\label{eq:My}
   M_{,y} &=& \frac{1-x^2}{2(x^2-y^2)}\Bigg[
    y(1-x^2)u_{,x}^{\ 2}+y(1-y^2)u_{,y}^{\ 2}-2x(1-y^2)u_{,x}u_{,y}\nonumber\\
   &&
   +4xy u_{,x}-2\frac{x^2+y^2-2x^2y^2}{1-x^2}u_{,y}-4y\nonumber\\
   &&
   +\frac{1-y^2}{1-x^2}\ee^{2u}
   \Big(y(1-x^2)Q_{,x}^{\ 2}+ y(1-y^2)Q_{,y}^{\ 2}
   -2x(1-y^2)Q_{,x}Q_{,y}\Big)\nonumber\\  
   &&
   +\frac{4\ee^u}{(1-x^2)R_0}\Big[\frac{1-x^2}{1-y^2}\ee^{-2u}
   \Big(y(1-x^2)A_{3,x}^{\ 2}+y(1-y^2)A_{3,y}^{\ 2}
    -2x(1-y^2)A_{3,x}A_{3,y}\Big)\nonumber\\
   &&
   +y(1-x^2)(A_{4,x}-QA_{3,x})^2+y(1-y^2)(A_{4,y}-QA_{3,y})^2\nonumber\\
   &&
   -2x(1-y^2)(A_{4,x}-QA_{3,x})(A_{4,y}-QA_{3,y})\Big]\Bigg].\label{eq:Meq2}
  \end{eqnarray}

Furthermore, Maxwell's equations lead to the following two equations,
\begin{equation}
   \left[\ee^u(A_{4,x}-QA_{3,x})\right]_{,x}
   =\left[\frac{1-y^2}{1-x^2}\,\ee^u(A_{4,y}-QA_{3,y})\right]_{,y},
\end{equation}
\begin{equation}
   \left[\frac{1-x^2}{1-y^2}\,\ee^{-u}A_{3,x}-Q\ee^u(A_{4,x}-QA_{3,x})\right]_{,x}
   =\left[\ee^{-u}A_{3,y}-\frac{1-y^2}{1-x^2}Q\,\ee^u(A_{4,y}-QA_{3,y})\right]_{,y}.
\end{equation}


\end{document}